\documentclass[a4paper]{article}
\usepackage{amsmath}
\usepackage{graphicx}

\begin{document}

\title{Opinion Dynamics Theory for Analysis of Consensus Formation and Division of Opinion on the Internet}

\author{\textbf{Akira Ishii}\\ Faculty of Engineering, Tottori University, Koayam, Tottori 680-8552, Japan\\  \textbf{Yasuko Kawahata}\\  Faculty of Social and Information Studies, \\
Gunma University, 4-2 Aramaki-machi, Maebashi,\\ Gunma, 371-8510, Japan
}

\maketitle

\begin{abstract}
    The massive amount of text data on the web has facilitated research on the quantitative analysis of public opinion, which could not be visualized earlier. In this paper, we propose a new opinion dynamics theory. This theory that is intended to explain agreement formation and opinion breakup division in opinion exchanges on social media such as Twitter. With the popularization of the public network, we have become able to communicate with instantaneity and interactivity beyond the temporal and spatial constraints．Research on quantitatively analyzing the distribution of opinion on public opinion that has not been visualized so far utilizing massive web text data is progressing．Our model is based on the Bounded Confidence Model, that expresses opinions in as continuous quantity values. However, in the Bounded Confidence Model, it was assumed that people with different opinions move not in disregard but ignoring opinions. Furthermore, in our theory, it modeled so that it can expresser model incorporates the influence from of the external pressure outside and the phenomenon depending on the surrounding situation.
    
\end{abstract}%

\section{Introduction}
Opinion dynamics has a long research history, and many studies have been conducted mainly in the field of sociology \cite{French1956,Harary1959,Abelson1964,DeGroot1974,Lehrer1975,Chatterjee1975,Chatterjee1977,Wagner1978}. Early studies assumed linearity; however, models incorporating nonlinearity have also been studied \cite{Krause1977,Bechmann1977,Hegselmann1998,Krause2000,Deffuant2000,Dittmer2001,Weisbuch2001}. Theoretical progress of the recent years on opinion dynamics are described in the review paper of S\^{i}rbu et. al.\cite{Sirbu2017} Consensus formation has been studied based on the local majority rule as an application of the renormalization group theory in physics \cite{Galam1999,Galam2000}. Moreover, applying the theory of magnetic physics, the theory for comparing opinion agreement and opposition to the direction of magnetic moment of magnetism has been studied in the field of social physics \cite{Galam1997,SznajdWeron2000,SznajdWeron2011}. 
Many mathematical theories on opinion dynamics treat opinions as discrete values of +1 and 0, or +1 and -1. In contrast, certain theories consider opinions as continuous numerical values that can change through the exchange of opinions with others. The bounded confidence model is a representative model of the theory that handles the continuous transition of opinion \cite{Hegselmann2002}.
In this study, we propose a theory that expresses opinions as continuous values and deals with changes in the opinion values due to the exchange of opinions with others. Moreover, we assume that the opinion of each people can be both positive or negative values. 
For example, in a study of Tweet on political situation in the United States, there is a study to classify political opinions from conservative to liberal by one-dimensional axis\cite{Bail2018}.
In this research, we assume that differences in opinion can be represented by one-dimensional axis values as in this reference. Based on this theory, it is possible to express the division of opinion in society, assuming that opinions of people who disagree with each other are exchanged, and the opinions of both are further divided. Such a division of opinions is a phenomenon often seen on social media such as Twitter. 

\section{Modelling opinion dynamics}
Our model is based on the original bounded confidence model of Hegselmann-Krause\cite{Hegselmann2002}. For a fixed agent, say $i$, where $1 \leq i \leq N$, we denote  the agent’s opinion at time $t$ by $I_i (t)$.  According to Hegselmann-Krause \cite{Hegselmann2002}, opinion formation of agent $i$ can be described as follows.

\begin{equation}
I_i(t+1) = \sum_{j=1}^N D_{ij} I_j(t)
\end{equation}

This can be written in the following form.

\begin{equation}
\Delta I_i(t) = \sum_{j=1}^N D_{ij} I_j(t) \Delta t
\end{equation}

where it is assumed that $D_{ij} \geq 0$ for all $i, j$ in the model of Hegselmann-Krause.  Using Based on this definition, $D_{ij}=0$  means that the opinion of agent $i$ is not affected by the opinion of agent $j$. 

 Here, as a result of exchanging opinions, consider the possibility that the opinions of two people with different opinions change move in different directions. Let's us think about consider the distribution of opinions with in the positive and negative directions of a one-dimensional axis. In this case, the value range of $I_i (t)$ is $- \infty \leq I_i (t) \leq + \infty$. If we get the continue value of $I_i(t)$ from -1 up to 1, we can introduce
 
 \begin{equation}
 Opinion_i (t) = tanh(I_i(t)).
 \end{equation}
 
 We modify the meaning of the coefficient $D_{ij}$ as the coefficient of trust. We assume here that $D_{ij}>0$ if there is a trust relationship between the two persons, and $D_{ij}<0$ if there is distrust relationship or consensus between the two persons. 
 
  Let $A(t)$ be the pressure at time $t$ from the outside and denote the reaction difference for each agent is denoted by the coefficient $c_i$. Therefore, the change in opinion of the agent can be expressed as follows.
  
\begin{equation}
\Delta I_i(t) = c_i A(t) \Delta t + \sum_{j=1}^N D_{ij} I_j(t) \Delta t
\end{equation}

We assume here that $D_{ij}$  and $D_{ji}$  are independent. Usually, $D_{ij}$ is an asymmetric matrix;  $D_{ij} \neq D_{ji}$. Moreover, $D_{ij}$  and $D_{ji}$  can have different signs.

Long-term behavior requires attenuation, which means that topics will be forgotten over time. Here we introduce exponential attenuation. The expression is as follows.

\begin{equation}
\Delta I_i(t) = -\alpha I_i(t) \Delta t + c_i A(t) \Delta t + \sum_{j=1}^N D_{ij} I_j(t) \Delta t
\end{equation}

\section{Opinion dynamics for two agents}

Let us first consider the case where the opinions of the two agents are the same. In the calculation below, we set $A(t)$ to be $Adv$ as a constant value. In the all calculations in this paper, we assume that $A(t)$ is constant  for simplicity in order to pay attention to the effect of $D_{ij}$. In the actual simulation of the real society behaviors, the external effect $A(t)$ is also significant and time-dependent.

\begin{figure}[h]
\begin{center}
\includegraphics[height=5cm]{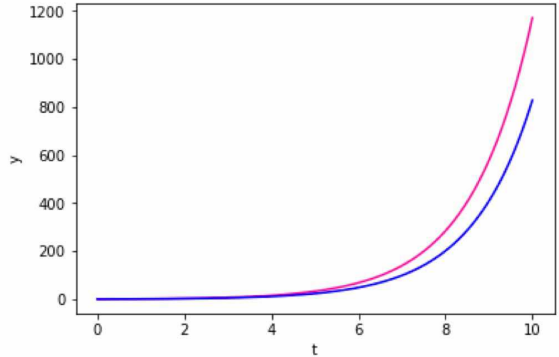}
\caption{Calculation result for N=2. Adv= 0.5, $D_{AB} = 1.0$, $D_{BA} = 0.5$. The initial value is $I_A(0)=0.005$, $I_B(0) = 0.2$.
}
\label{N=2a}
\end{center}
\end{figure}

In this case, both opinions are positive. If  $D_{ij}>0$, $D_{ij}$ and  $I_j (t)$ is positive. Thus, the opinion $I_i (t)$ moves in the positive direction as shown in fig.\ref{N=2a}. This means that by having a conversation with an agent of the same positive opinion, agent $i$ will change its opinion to be more and more positive. Similarly, if the opinions of both agents are the same negative opinion, the opinions become more and more negative.

\begin{figure}[h]
\begin{center}
\includegraphics[height=5cm]{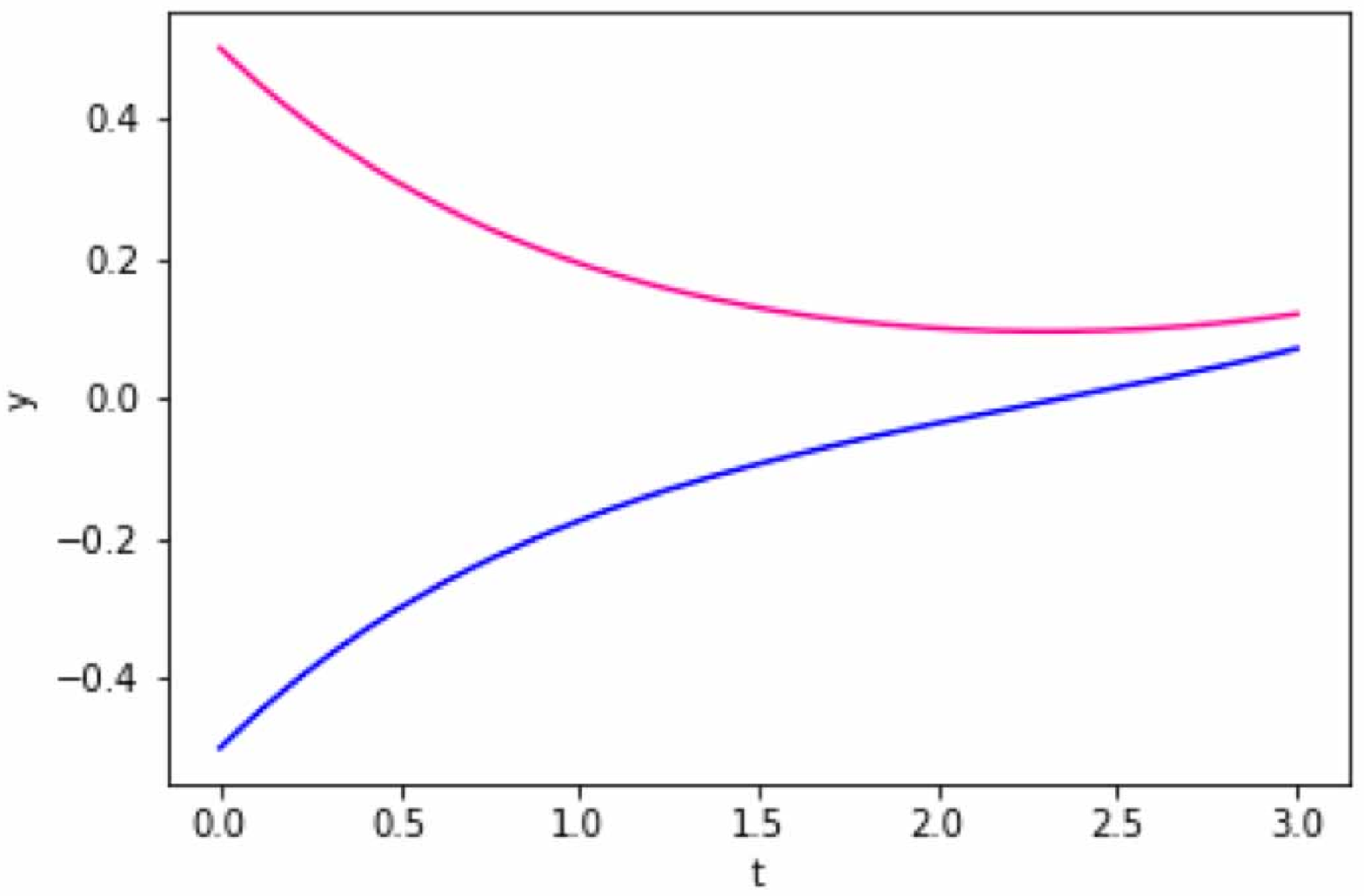}
\caption{Calculation result for N=2. Adv= 0.005, $D_{AB} = 1.0$, $D_{BA} = 1.0$. The initial value is $I_A(0)=0.5$, $I_B(0) = -0.5$.}
\label{N=2b}
\end{center}
\end{figure}

In Fig.\ref{N=2b}, A is a positive opinion and B is a negative opinion, but it is a case where they trust each other. Namely, we consider the case where the opinions of the two agents are opposite: $I_A (t)>0$ and $I_B (t)<0$ where  both $D_{AB}$ and $D_{BA}$ are positive values. In this case, $D_{AB} I_B (t)$ is negative. Thus, the opinion of agent $A$ moves to negative because of the effect of agent $B$ having the opposite opinion. The reason why the opinion of agent $A$ moves negatively is that there is a trust relationship with agent $B$. For the case of $I_A (t)<0$ and $I_B (t)>0$ with  $D_{AB}>0$, the result is the same.

\begin{figure}[h]
\begin{center}
\includegraphics[height=5cm]{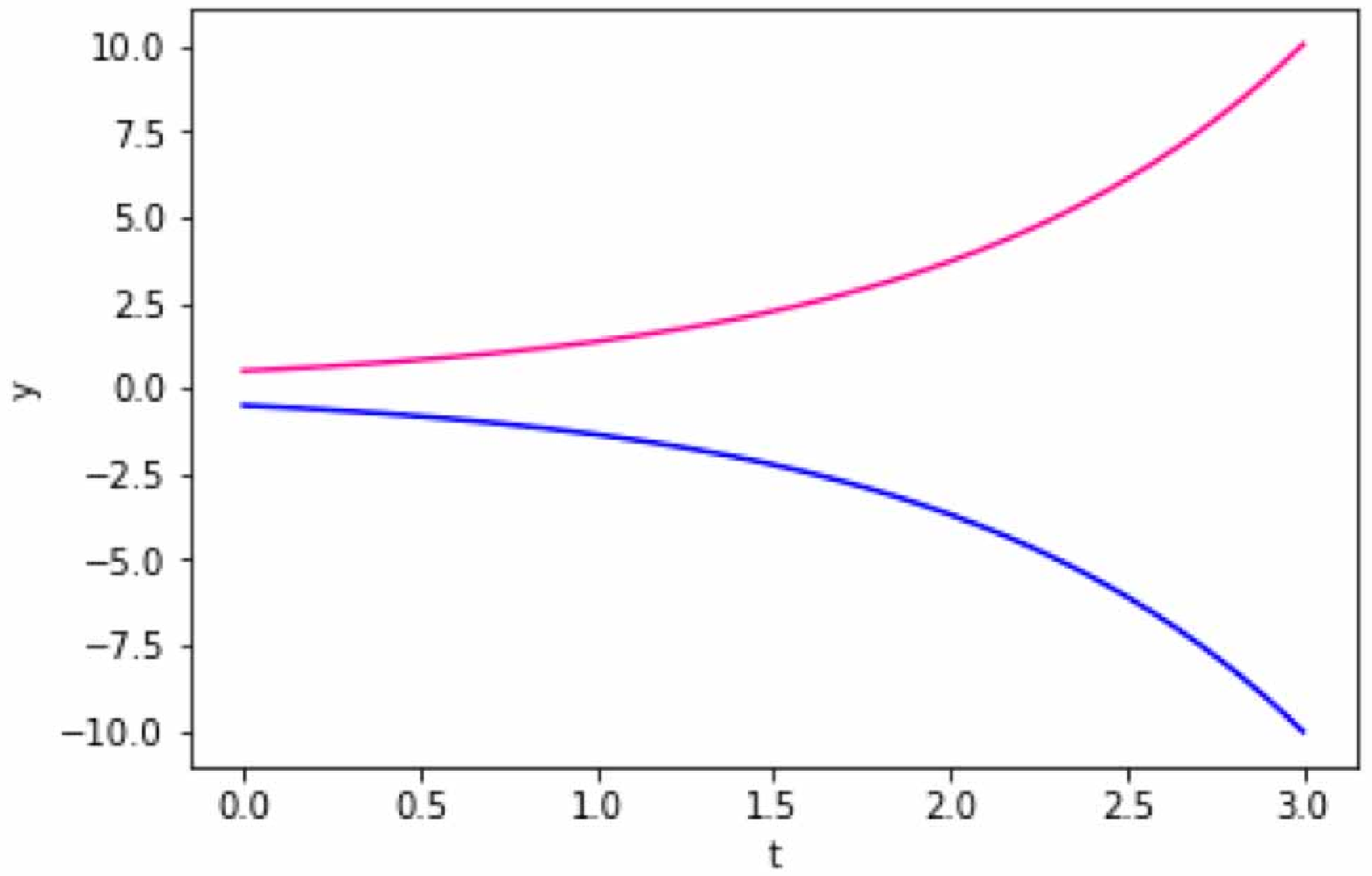}
\caption{Calculation result for N=2. Adv= 0.005, $D_{AB} = -1.0$, $D_{BA} = -1.0$. The initial value is $I_A(0)=0.5$, $I_B(0) = -0.5$.}
\label{N=2c}
\end{center}
\end{figure}

Next, we consider the case where the opinions of the two agents are opposite：$I_A (t)>0$ and $I_B (t)<0$ where  $D_{AB}<0$. The calculated result is shown in Fig.\ref{N=2c}. In this case, $D_{AB} I_B (t)$ is positive. Thus, the opinion of agent A moves to more positive. It means that, in a discussion with an agent in disagreement, as there is no trust relationship with that agent, we consider that agent A held his/her opinion more firmly. Similarly, the agent B held his/her opinion more firmly, too. For the case of $I_A (t)<0$ and $I_B (t)>0$ with  $D_{AB}<0$, the result is same. 
This result shows that the dialogue of people who do not trust each other never leads to an agreement.

\section{Opinion dynamics for three agents}

Next, calculations in the case of three people are shown. A has a positive opinion, B has a negative opinion, and a third person C has an almost neutral opinion. 
If C's opinion is zero, it obviously does not affect A or B. Even if it is neutral, it is important that C have an opinion on which side. In the following it is assumed to have a slightly positive opinion.

\begin{figure}[h]
\begin{center}
\includegraphics[height=5cm]{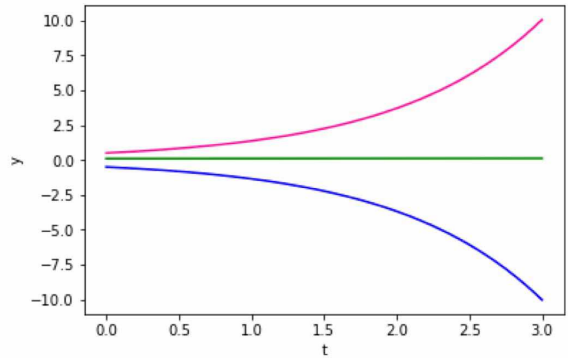}
\caption{Calculation result for N=3. Adv = 0.005,
    $D_{AB} = -1.0$,
    $D_{AC} = 0.1$,
    $D_{BA} = -1.0$,
    $D_{BC} = 0.1$,
    $D_{CA} = 0.1$,
    $D_{CB} = 0.1$.
The initial value is $I_A(0)=0.5$, $I_B(0) = -0.5$, $I_C(0) = 0.1$.    }
\label{N=3a}
\end{center}
\end{figure}

In a situation similar to Fig.\ref{N=2c}, C is almost neutral in opinion, not much trusted by A and B, and $D_{AC}$ and $D_{BC}$ are small values. The calculation result is shown in Fig.\ref{N=3a}, but it is almost the same as Fig.\ref{N=2c}, and A and B repel each other. That is, the influence of C is very small in this case.

\begin{figure}[h]
\begin{center}
\includegraphics[height=5cm]{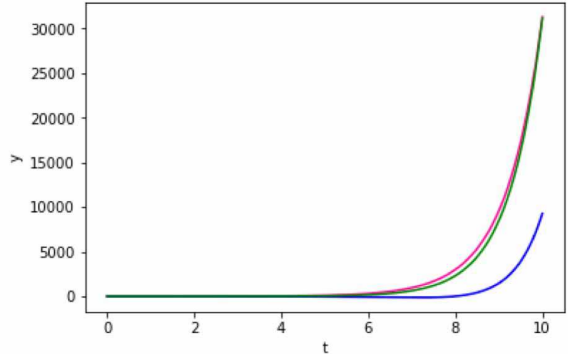}
\caption{Calculation result for N=3. Adv = 0.005,
    $D_{AB} = -1.0$,
    $D_{AC} = 1.5$,
    $D_{BA} = -1.0$,
    $D_{BC} = 1.5$,
    $D_{CA} = 1.0$,
    $D_{CB} = 1.0$.
The initial value is $I_A(0)=0.5$, $I_B(0) = -0.5$, $I_C(0) = 0.1$. $t=0$ to 10.}
\label{N=3b}
\end{center}
\end{figure}

\begin{figure}[h]
\begin{center}
\includegraphics[height=5cm]{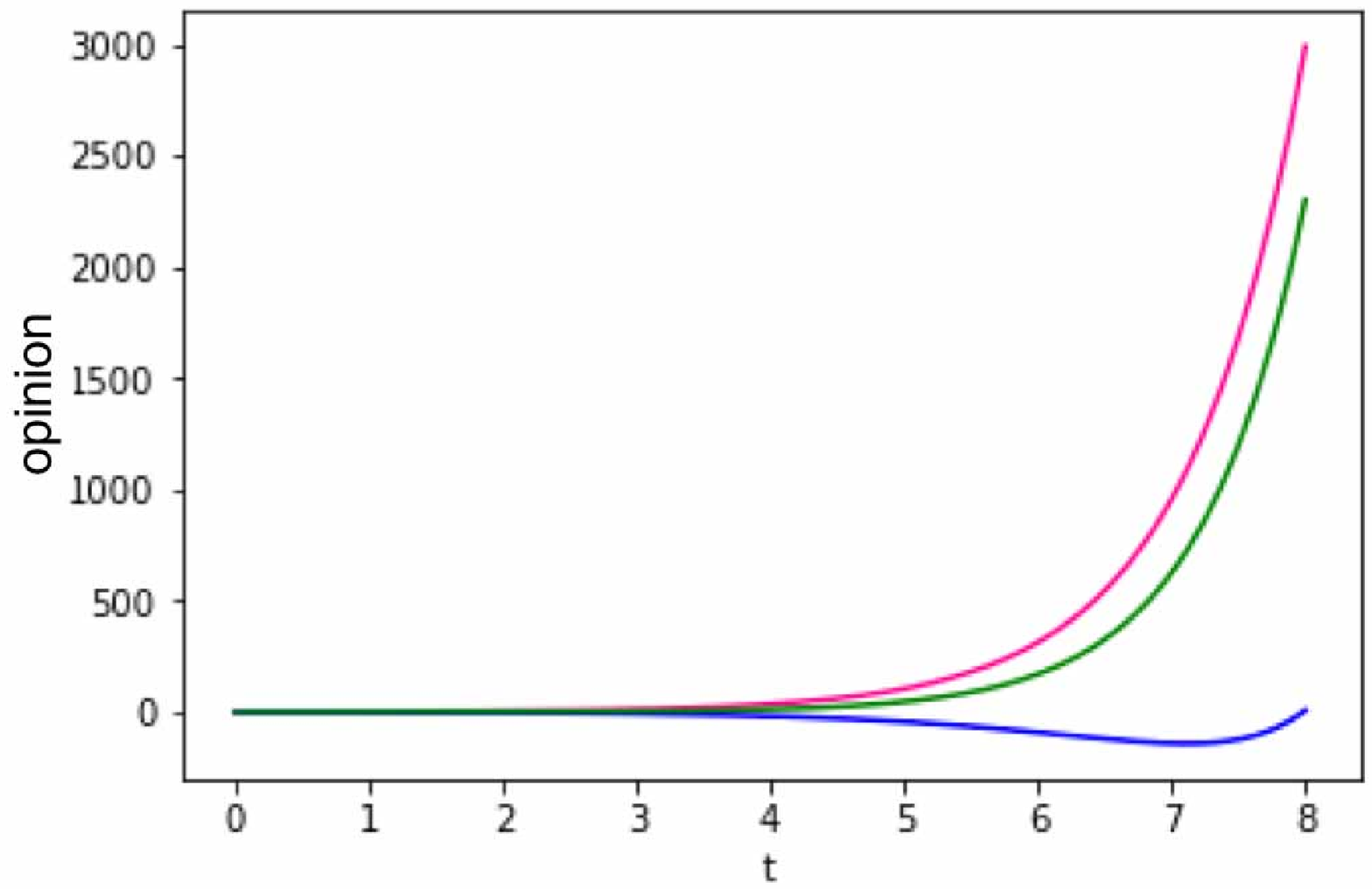}
\caption{Calculation result for N=3. Adv = 0.005,
    $D_{AB} = -1.0$,
    $D_{AC} = 1.5$,
    $D_{BA} = -1.0$,
    $D_{BC} = 1.5$,
    $D_{CA} = 1.0$,
    $D_{CB} = 1.0$.
The initial value is $I_A(0)=0.5$, $I_B(0) = -0.5$, $I_C(0) = 0.1$.  $t=0$ to 8.}
\label{N=3c}
\end{center}
\end{figure}

However, if C is strongly trusted by both A and B, the situation will be different. Here we consider that A is a positive opinion and B has a negative opinion and they do not trust each other. However, both have strong confidence in C. Then, as shown in Fig.\ref{N=3b} and Fig.\ref{N=3c}, the opinions of A and B work in the direction converging with C as a vector. In this case, the initial value of opinion of C is a weak positive opinion, but if it is a weak negative opinion, the opinions of A and B are gathered in a negative direction.

A and B's opinion are getting closer by brokerage of C who is trusted. Approach to C's opinion. In other words, C is a mediator with strong political power that can solve conflict. One example of the person C would be the former president of the Republic of South Africa, Nelson Rolihlahla Mandela who instructed the Republic of South Africa so that all peacefully settled.

\section{Discussion}

From a simple result of N = 2 and N = 3, we can guess to some extent what kind of social phenomena this theory can explain even in the case of many people. The basic equation to solve many person problem is 

\begin{equation}
\Delta I_i(t) = -\alpha I_i(t) \Delta t + c_i A(t) \Delta t + \sum_{j=1}^N D_{ij} I_j(t) \Delta t
\end{equation}

The programming of this equation with general N person is very easy.

\begin{figure}[h]
\begin{center}
\includegraphics[height=5cm]{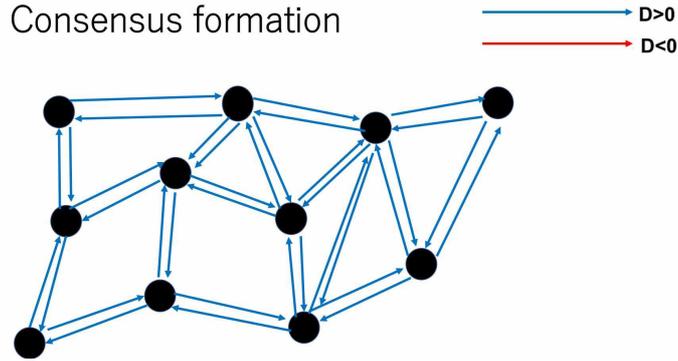}
\caption{Schematic illustration of consensus formation of this theory.}
\label{consensus}
\end{center}
\end{figure}

In Fig.\ref{consensus}, there is the most ideal social setting. If all N people trust each other strongly, society can be guessed that conflict disappears, opinion of people is the same, problem does not occur. 
As you can see immediately, this situation can easily be calculated with the mathematical model of this study.
It is ideal, so to speak, it is similar to the world of song Imagine of John Lennon. It is not realized in actual society. But if the village where a small number of people live is relatively isolated, it would be a possible setting.

     Suppose a group of N people is divided into two: a group with positive opinions and a group with negative opinions. All members of the group with positive opinions are connected by trust, and all members of the group with negative opinions are also connected by trust. In this case, consensus building within each group occurs smoothly. However, assume that there is no trust relationship between the group with positive opinions and the group with negative opinions. Therefore, $D_{ij} <0$ between these two groups. 
     
\begin{figure}[h]
\begin{center}
\includegraphics[height=5cm]{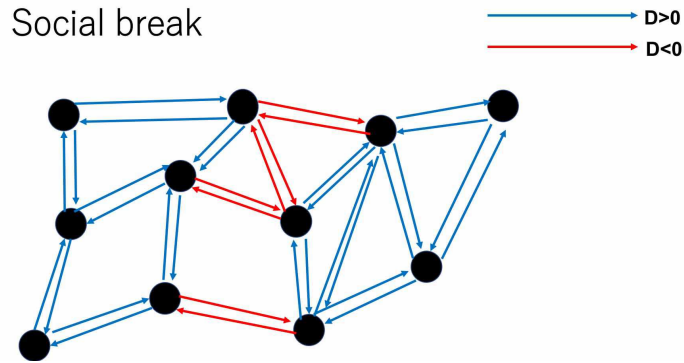}
\caption{Schematic illustration of social break of this theory.}
\label{socialbreak}
\end{center}
\end{figure}

     The conflicting opinions are promoted because the group with positive opinions and the group with  negative opinions have no mutual trust relationship. Consequently, it leads to the division of society. The typical example is ecology people and anti-ecology people. In the political history of Japan, at the time of the Meiji Restoration turmoil, the factions wanting to preserve the old regime and the factions establishing a new government fought violently, and this was finally resolved in the civil war of 1868 in Japan. This is a typical example of society being divided. It is believed that a social consensus did not result because the two sides fought intensely and there was no trust relationship.
The schematic illustration of this social break is shown in fig.\ref{socialbreak}.

Here, we consider a person isolated in a group. The opinion of the group is positive, whereas only the opinion of the isolated person is negative. As a group consisting of $N$ people, $N-1$ persons excluding the isolated person do not trust the isolated person, while that isolated person also does not trust the other $N-1$ persons. In this case, $D_{ij} <0$ between the isolated person and the other persons in the group. In this case, the opinions of the $N-1$ people in the group become increasingly positive, and the opinion of the isolated person becomes increasingly negative. Therefore, an isolated person and another member of the group are completely divided.

\begin{figure}[h]
\begin{center}
\includegraphics[height=5cm]{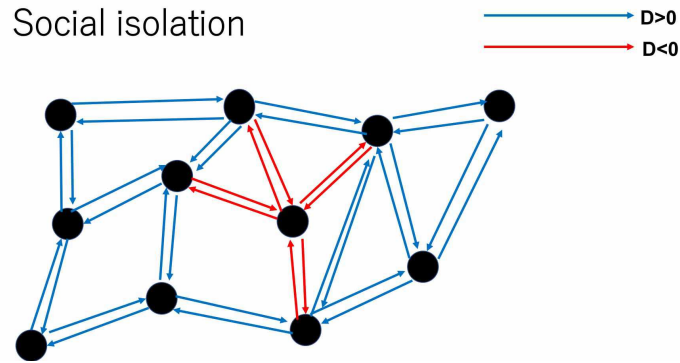}
\caption{Schematic illustration of social isolation of this theory.}
\label{Isolation}
\end{center}
\end{figure}

      Even if $N-1$ people in the group do not trust an isolated person, the situation would be different if the isolated person trusts the other $N-1$ persons in the group. In this case, let $i$ be an isolated person, $D_{ij}> 0$ and $D_{ij} <0$. The opinion of the isolated person gradually approaches the group's opinion owing to the influence of the other people in the group and external forces such as media, such that isolation is gradually resolved. Moreover, if other people in the group change to trust the people who are isolated, the isolation will be resolved more quickly. Whether this situation can be solved, for example, by trusting this person by a small number of people can be calculated by the mathematical model of this research.
The schematic illustration of this social isolation is shown in Fig.\ref{Isolation}.

     If an individual is afraid of isolation, it is considered that the social group or society as a whole cannot exclude individuals for opinion of members. This fear of isolation is a case study that often results in silence rather than expressing an opinion. Media is an important factor related to both dominant ideas and people's perception of dominant ideas, and the assessment of the social environment does not necessarily correlate with reality. 
     
     People fear social isolation, and try to avoid isolation as $D_{ij}> 0$ among them. Moreover, when trying to enter the majority in society, each individual becomes confused as to which opinion is the majority. Based on mass media information and the local majority of surrounding people, we estimate which opinion is the majority opinion. In our new opinion dynamics theory, the effect of mass media can be included by $A(t)$ . 
     
     As a result of combining these effects, the majority is formed rapidly. Even in the opinion of the majority in the early stage, if opinions are contrary to the information from mass media and surrounding people, people with different opinions become silent as $D_{ij}> 0$ within the surroundings. Thus, the silence spiral \cite{Noelle-Neumann1974}  can also be explained by our new opinion dynamics theory.

In future calculations, the external effect $A(t)$ as mass media information or statement from government will also play an important role in the simulation using our new opinion dynamics theory of this paper. In that case, the external effect $A(t)$ is the time-dependent function. Moreover, the coefficient $c_i$ can be positive or negative value. In the case of negative $c_i$, the person $i$ obtain negative effect from the external effect $A(t)$.

\section{Conclusion}

In this research, we presented a theory of opinion dynamics that considers the opinion of each person a continuous value, rather than a discrete value. Opinions are represented by real numbers ranging from positive to negative. We introduce "trust" and "distrust" as a coefficient of each person pairs. In addition to the influence of opinion exchanges within each group, we constructed a mathematical model that incorporates external pressure. Using this theory, we can mathematically express many phenomena that can occur in a group in society. 

In this new opinion dynamics theory, it is possible to calculate the dynamics of a complicated system mixed with people's trust and suspicion. Also, as there is no upper limit on the opinion, we can explain the situation where opinions are getting sharper and sharper. Simulation of a large number of people is also prepared. In the future, we will compare and examine which case is assumed whether
this theory conforms to actual data concerning speech in
actual political and social problems.

As a future prospect, we would like to conduct researches on reactions to various social risks, discussions on Fake News. Definition differs depending on the scope such as macro disposers, civil disputes, shocks of market transactions etc. Generally occurring in a microscopic range such as disease, crime, ethnic discrimination.

\section*{Aknowlegement}

The authors are grateful for discussion with Prof. Serge Galam.

\end{document}